\documentclass{ws-mpla}
\usepackage{epsf}

\newcommand{\nn}{\nonumber}
\newcommand{\be}{\begin{equation}}
\newcommand{\ee}{\end{equation}}
\newcommand{\ba}{\begin{eqnarray}}
\newcommand{\ea}{\end{eqnarray}}
\newcommand{\req}[1]{(\ref{#1})}
\def\={\,=\,}
\newcommand{\ci}[1]{\cite{#1}}

\def\gev{~{\rm GeV}}

\def\ev{~{\rm eV}}
\def\ale{\alpha_{\rm elm}}
\def\als{\alpha_{\rm s}}
\def\et1{\widetilde{\eta}_1}
\def\et8{\widetilde{\eta}_8}
\def\eps{\epsilon}

\def\etp{\eta^\prime}

\def\ab{\bar{a}}
\def\ub{\bar{u}}
\def\db{\bar{d}}
\def\sb{\bar{s}}
\def\cb{\bar{c}}

\begin{document} 
\markboth{Kroll}{isospin}

\catchline{}{}{}{}{}

\title{\bf ISOSPIN SYMMETRY BREAKING \\[0.3em]
 THROUGH $\pi^0$ - $\eta$ - $\etp$ MIXING}

\author{\footnotesize P.\ KROLL} 

\address{Fachbereich Physik, Universit\"at Wuppertal,\\ 
D-42097 Wuppertal, Germany\\
kroll@physik.uni-wuppertal.de}
\maketitle

\begin{abstract}
Mixing of the pseudoscalar mesons is discussed in the quark-flavor
basis. The divergences of the axial vector currents which embody 
the axial vector anomaly, combined with the assumption that the decay
constants follow the pattern of particle state mixing in that basis,
determine the mixing parameters for given masses of the physical
mesons. These mixing parameters are compared with results from other 
work in some detail. Phenomenological applications of the 
quark-flavor mixing scheme are presented with particular interest
focussed on isospin symmetry breaking which is generated by means of  
$\eta$ and $\etp$ admixtures to the pion. Consequences of a possible 
difference in the basis decay constants $f_u$ and $f_d$ for the
strength of isospin symmetry breaking are also examined.
   
\end{abstract}
\ccode{PACS Nos.: 12.38.Aw, 13.25.Gv, 14.40.Aq}

\section{Introduction}
That the character of the approximate flavour symmetry of QCD is
determined by the pattern of the quark masses is a well-known fact 
that has been extensively discussed in the literature for decades. 
Isospin symmetry in particular which would be exact for identical $u$
and $d$ quark masses, $m_u=m_d$, holds to a rather high degree 
of accuracy empirically, although the ratio $(m_d-m_u)/(m_d+m_u)$  is
about 1/3 and not, as one would expect for a true symmetry, much
smaller than unity. Violations of isospin symmetry
for pseudoscalar mesons within QCD are generated by admixtures 
of the $\eta$ and $\etp$ to the pion. On exploiting 
the divergences of the axial vector currents but neglecting $\eta
-\etp$ mixing, Gross, Treiman and Wilczek~\ci{gross} obtained for the
mixing angle between the pion and the flavor-octet state $\eta_8$ 
\be
\eps_{\,0}\=\frac{\sqrt{3}}{4}\, \frac{m_d-m_u}{m_s - (m_u+m_d)/2}\,,
\label{pcac}
\ee 
a result that also follows from lowest order chiral perturbation
theory~\ci{gasser}. We learn from \req{pcac} that due to the effect of
the U$_{\rm A}$(1) anomaly which is embodied in the divergences of the
axial vector currents, isospin symmetry breaking (ISB) is of the order 
of $(m_d-m_u)/m_s$ instead of the expected $(m_d-m_u)/(m_d+m_u)$. 
Isospin symmetry is thus partially restored and amounts to only a few
percent for pseudoscalar mesons. It is therefore to be interpreted  
rather as an accidental symmetry which comes about as a consequence of the
dynamics. For hadrons other than pseudoscalar mesons the strength
of ISB is not necessarily set by the mass ratio \req{pcac}; for comments
on ISB in the vector meson sector see Ref.~\refcite{fritzsch}.

Due to a number of recent experiments the interest in ISB has been
renewed. It therefore seems opportune to review recent progress in our
understanding of $\pi^0 - \eta - \etp$ mixing. In an analysis of 
$\eta-\etp$ mixing~\ci{FKS1} the quark-flavor basis has been used and 
assumed that the decay constants in that basis follow the pattern of
particle state mixing. With the help of the divergences of the axial 
vector currents the basic parameters of that mixing scheme can be
determined for given masses of the physical mesons. It has been found
in Refs.~\refcite{FKS1} and \refcite{FKS2} that this approach leads to 
consistent results and explains many empirical features of $\eta-\etp$ 
mixing. This mixing scheme will be presented in Sect.\ 2 and, in 
Sect.\ 3, the phenomenology of $\eta - \etp$ mixing briefly reviewed. 
Sect.\ 4 is devoted to a discussion of mixing in the flavor 
octet-singlet basis and to a detailed comparison with other recent
work on $\eta-\etp$ mixing.  

The inclusion of the $\pi^0$ in the quark-flavor mixing scheme is
straightforward~\cite{FKS2}. Here, in this work,
$\pi^0-\eta-\etp$ mixing will be discussed in great detail and,
as a new aspect, the role of possible differences in the decay
constants $f_u$ and $f_d$ will be investigated. It is important to
realize that the decay constants which represent wavefunctions at zero
spatial quark-antiquark separation, are functions of the quark
masses. Hence, ISB generated through $f_u\neq f_d$ is related to the
$u - d$ quark-mass difference, too. It is only due to our
inability of calculating the decay constants to a sufficient
degree of accuracy at present that we have to consider them as
independent soft parameters. ISB will be discussed in Sect.\ 5. 
The vacuum-$\pi^0$ matrix element of the topological charge density is
calculated in Sect.\ 6 and phenomenological consequences are
presented. Finally, in Sect.\ 7, isospin symmetry violating processes, 
which are not directly controlled by the U$_{\rm A}$(1) anomaly, are
examined. The conclusions are provided in Sect.\ 8.    

\section{Mixing in the quark-flavor basis} 
The quark-flavor basis is constructed by means of the states $\eta_{\,a}$, $a=u,d,s$ 
which are understood to possess the parton compositions
\be
|\eta_{\,a}\rangle \= \Psi_a \, |a\ab \rangle + \cdots\,
\label{def-Fock}
\ee
in a Fock expansion. Here $\Psi_a$ denotes a (light-cone) wavefunction. 
The higher Fock states, whose presence are indicated by the ellipses,
also include two-gluon components. Because of the fact that light-cone
wave functions do not depend on the hadron's momentum, one can define
decay constants
\be
f_a \= 2\sqrt{6}\, \int \frac{dx\, d^2k_\perp}{16\pi^3}\,
\Psi_a(x,k_\perp)\,,
\label{def-dec}
\ee
associated with the states $\eta_a$. The variable $x$ denotes the
usual (light-cone plus) momentum fraction the quark carries and $k_\perp$
its transverse momentum with respect to its parent meson's
momentum. The definition \req{def-dec} is exact, only the $q\bar{q}$
component contributes to the decay constants. The basic decay
constants are assumed to possess the property 
\be
\langle0\mid J^a_{5\mu}|\eta_{\,b}(p)\rangle \= \delta_{ab}\, f_a\, p_\mu\,
\qquad a,b=u,d,s\,,
\label{currents}
\ee
where $J^a_{5\mu}=\ab \gamma_\mu \gamma_5\, a$ denotes the
axial-vector current for quarks of flavor $a$. Eqs. \req{def-Fock},
\req{def-dec} and \req{currents} form the basis of the quark-flavor
mixing scheme proposed in Ref.~\refcite{FKS1}. This mixing scheme
holds to the extent that Okubo-Zweig-Iizuka (OZI)-rule violations,
except those mediated by the U$_A$(1) anomaly enhanced
$q_i\bar{q}_i\to q_j\bar{q}_j$ transitions, can be neglected~\footnote{
OZI-rule violations are of order $1/N_c$ and become negligible in the
large $N_c$ limit.}. Flavor symmetry
breaking, on the other hand, is large and is to be taken into account
in any mixing scheme for the pseudoscalar mesons.  

Since mixing of the $\pi^0$ with the $\eta$ and $\etp$ is weak while
$\eta-\etp$ mixing is strong it is appropriate to employ isoscalar and 
isovector combinations 
\be
\eta_{\,\pm} \= \frac1{\sqrt{2}}\, [\eta_{\,u}\pm \eta_{\,d}] \,,
\ee
as basis states instead of $\eta_u$ and $\eta_d$. The unitary
matrix $U$ that transforms from the basis $\{\eta_{\,-},
\,\eta_{\,+},\,\eta_{\,s}\}$ to the physical meson states 
$\{P_1=\pi^0,\, P_2=\eta,\, P_3=\etp\}$ can then be linearized in the
$\pi^0-\eta$ and $\pi^0-\etp$ mixing angles. An appropriate
parameterization of $U$ reads ($\beta,\psi\propto {\cal O}(\lambda)$,
$\lambda\ll 1$)
\be 
U(\phi,\beta,\psi) \=\left(\begin{array}{ccc}
          1 & \beta +\psi \cos{\phi} & -\psi \sin{\phi} \\
       -\psi -\beta \cos{\phi} & \cos{\phi} & -\sin{\phi}\\ 
       -\beta \sin{\phi} & \sin{\phi} & \phantom{-}\cos{\phi} 
                                \end{array} \right)\,,
\label{matrix}
\ee
with $U\,U^{\dagger}=1+{\cal O}(\lambda^2)$. The $\eta - \etp$ sector of
$U$ is identical to the parameterization utilized in Ref.~\refcite{FKS1}.
It is also of advantage to introduce isoscalar and isovector axial
vector currents   
\be
J^{\mp}_{5\mu}\= \frac1{\sqrt{2}}\, [\ub\gamma_\mu \gamma_5\, u \mp \db\gamma_\mu
    \gamma_5\, d]\,.
\ee
The matrix elements $\langle 0|J^b_{5\mu}|\eta_{b'}\rangle$
($b,b'=-,+,s$) then define new decay constants which can be collected 
in a decay matrix 
\be
{\cal F} \=  \left( \begin{array}{ccc}
       f_+ & z f_+ & 0 \\
       z f_+ & f_+ & 0  \\
       0 & 0 & f_s \end{array} \right)\,,
\label{decay}
\ee
In contrast to \req{currents}, ${\cal F}$ is non-diagonal. The decay
constants $f_+=(f_u+f_d)/2$ and $f_s$  are the basic decay constants
in the $\eta$-$\etp$ sector while the parameter $z=(f_u-f_d)/(f_u+f_d)$
occurs in $\pi^0 - \eta, \eta^\prime$ mixing; it is obviously of order 
$\lambda$. The decay constants are in principle renormalization 
scale dependent \ci{kaiser}. Ratios of decay constants or mixing
angles are, on the other hand, scale independent. Since the anomalous 
dimension controlling the scale dependence of the decay constants is
of order $\als^2$, this effect is tiny and discarded here. This
simplification is consistent with the neglect of OZI-rule violations~\ci{feld}. 

Taking vacuum-particle matrix elements of the current divergences,
one finds with the help of \req{currents}, \req{matrix} and \req{decay}
($a,a'=1,2,3$; $b, b'= -,+,s$) 
\be
\langle 0|\partial^\mu J^b_{5\mu}|P_a\rangle \= {\cal
  M}^2_{aa'}\,U_{a'b'}\, {\cal F}_{b'b}\,,
\label{divergence}
\ee
where ${\cal M}^2={\rm diag}{[M_{\pi^0}^2, M_\eta^2, M_{\etp}^2]}$ is the
particle mass matrix which appears necessarily quadratic here.
Next, I recall the operator relation~\cite{witten}
\be
\partial^\mu\, J^a_{5\mu} \= 2 m_a \ab\, i\gamma_5\, a + \omega\,,
\label{anomaly}
\ee
which holds as a consequence of the U$_{\rm A}$(1) anomaly.
The topological charge density is given by $\omega=\als/4\pi
G\tilde{G}$ where $G$ denotes the gluon field strength tensor and
$\tilde{G}$ its dual. Inserting \req{anomaly} into \req{divergence}
and neglecting terms of order $\lambda^2$, one obtains a set of 
equations which can be solved for the mixing parameters 
\ba
\sin{\phi} &=& \sqrt{\frac{(M_{\etp}^2-m_{ss}^2)(M_\eta^2-M_{\pi^0}^2)}
                 {(M_{\etp}^2-M_\eta^2)(m_{ss}^2-M_{\pi^0}^2)}}\,,\nn\\[0.2em]
a^2&=&\frac{\sqrt{2}}{f_+}\,\langle 0|\omega|\eta_+\rangle \=  
                       \frac{(M_{\etp}^2-M_{\pi^0}^2)(M_\eta^2-M_{\pi^0}^2)}
                        {(m_{ss}^2-M_{\pi^0}^2)} \,, \nn\\[0.2em]
y &=& \sqrt{2\, \frac{(M_{\etp}^2-m_{ss}^2)(m_{ss}^2-M_\eta^2)}
                     {(M_{\etp}^2-M_{\pi^0}^2)(M_\eta^2-M_{\pi^0}^2)}}\,,
\label{mix-rel}
\ea 
and 
\ba
\beta&=& \frac12\,\frac{m^2_{dd}-m_{uu}^2}{M_{\etp}^2 - M^2_{\pi^0}} +z\,,\nn\\[0.2em]
\psi &=& \frac12\,\cos{\phi}\, \frac{M_{\etp}^2 - M^2_{\eta}}{M_{\etp}^2 - M^2_{\pi^0}}\,
                         \frac{m^2_{dd}-m^2_{uu}}{M_\eta^2 - M^2_{\pi^0}}\,.
\label{iso-rel}
\ea
The pion's mass and decay constant fix to more parameters
\be
\frac12\, (m_{uu}^2 + m_{dd}^2)\= M_{\pi^0}^2\,, \qquad f_+\=f_\pi\,,
\ee
up to corrections of order $\lambda^2$.
Finally, the symmetry of the mass matrix forces relations between the decay
constants and the matrix elements of the topological charge density
\be
y\=\frac{f_+}{f_s}\=\sqrt{2}\,\frac{\langle 0|\omega|\eta_{\,s}\rangle}
            {\langle 0|\omega|\eta_{\,+}\rangle}\,, \qquad
z\=\frac{f_u-f_d}{f_u+f_d}\= - \frac{\langle 0|\omega|\eta_{\,-}\rangle}
            {\langle 0|\omega|\eta_{\,+}\rangle}\,.
\label{eq:yz}
\ee
The quark mass terms in the above equations are defined as matrix
elements of the pseudoscalar currents 
\be 
m^2_{aa} \=  \langle 0|2\, \frac{m_a}{f_a}\,  \ab\, i\gamma_5\, a|\eta_{\,a}\rangle\,.
\ee

The quark-flavor mixing scheme can readily be extended to the case of the
$\eta_c$~\cite{FKS1}.
\section{$\eta-\etp$ mixing}
\begin{figure}
\begin{minipage}{0.45\textwidth}
\includegraphics[width=0.90\textwidth,bb=130 370 415 498,clip=true]
                {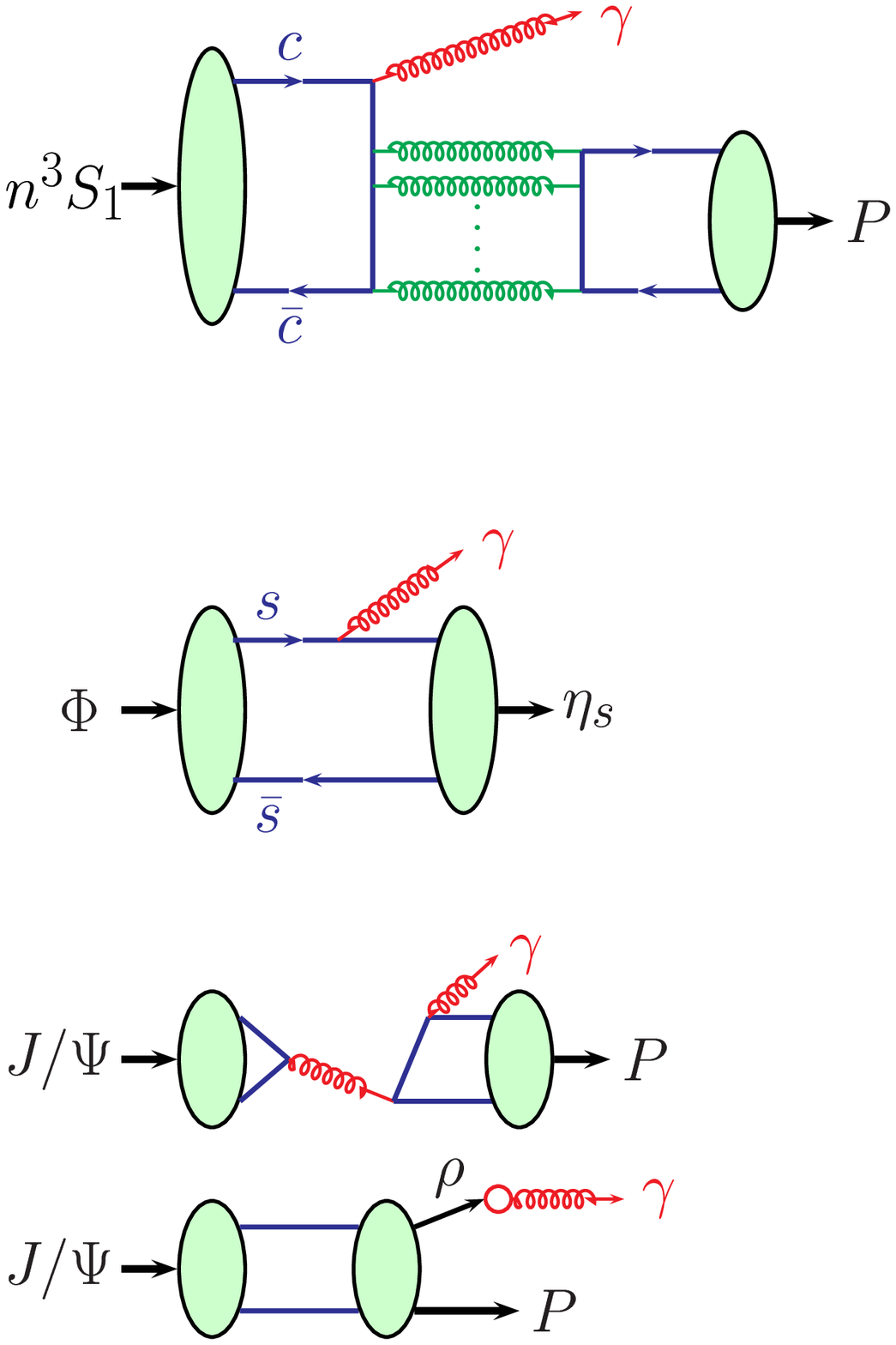}
\end{minipage}
\begin{minipage}{0.49\textwidth}
\includegraphics[width=0.95\textwidth, bb=109 570 471 710,clip=true] 
{fig-decay.ps}
\end{minipage}
\caption{Graphs for the radiative decays of the $\phi$
  meson (left) and the $n^3 S_1$ heavy quarkonium state (right).} 
\label{fig-decays}
\end{figure}

The three relations \req{mix-rel}, taken from Ref.~\refcite{uppsala}, fix 
the $\eta-\etp$ mixing parameters for given
masses of the physical mesons if the current algebra result
\be
m_{ss}^2\= 2 M^2_{K^0} - M^2_{\pi^0}\,,
\label{constraint}
\ee
is adopted for the strange quark mass term. This theoretical
estimate of the $\eta, \etp$ mixing parameters~\footnote
{Note that the theoretical estimate presented
here differs slightly from the one presented in Ref.~\refcite{FKS1} 
where an additional constraint on $f_s$ has been employed.}
provides the values quoted in Tab.\ \ref{tab:1}.
\begin{table}[h]
\renewcommand{\arraystretch}{1.3}
\tbl{Theoretical and phenomenological $\eta - \etp$ mixing parameters
  in the quark-flavor mixing scheme~\protect\ci{FKS1}.} 
{\begin{tabular}{l|ccccc} 
      &$\phi$ & $y$ &$a^2 [\gev^2]$ &$f_+/f_{\pi}$& $f_s/f_{\pi}$\\ \hline
theory& $41.4^\circ$& 0.79 & 0.276 & 1.00 & 1.27 \\
phenom& $39.3^\circ$& 0.81 & 0.265 & 1.07 &1.34  \\ 
      & $\pm 1.0^\circ$ & $\pm 0.03$ & $\pm 0.010$ & $\pm 0.02$ & $\pm 0.06$\\ \hline
\end{tabular}}
\label{tab:1}
\end{table} 
\renewcommand{\arraystretch}{1}
In order to take full account of flavor symmetry breaking a
phenomenological determination of the mixing parameters has also been
attempted in Ref.~\refcite{FKS1} by using experimental data instead of
the theoretical result \req{constraint}. This analysis includes a
number of processes like radiative transitions between light vector 
($V$) and pseudoscalar ($P$) mesons or scattering processes like 
$\pi^- p \to Pn$, which all rely on the validity of the OZI rule. The 
$\eta/\etp$ ratios of corresponding processes provide the mixing angle
$\phi$. Other dynamical effects are in general expected to cancel in the ratios, 
only phase space corrections have to be considered. An example is set 
by the radiative decays of the $\phi$ meson. These decays proceed
through the emission of the photon from the strange quark and a
subsequent $s\bar{s}$ transition into the $\eta_s$ component, see
Fig.\ \ref{fig-decays}. Hence, the ratio of $\eta$ and $\etp$ decay
widths reads
\be
\frac{\Gamma[\phi\to \gamma \etp]}{\Gamma[\phi\to \gamma \eta\,]} \=
\cot^2{\phi} \left(\frac{k_{[\phi\,\gamma\,\etp]}}
                          {k_{[\phi\,\gamma\,\eta]}}\right)^3
\,\Big[1- 2\frac{f_s}{f_+}\,\frac{\tan{\phi_V}}{\sin{2\phi}}\Big]\,,
\label{phi-ratio}
\ee
where
\be
k_{[\alpha\,\beta\,\gamma]}\=\frac12 M_\alpha\,
           \sqrt{1-2\frac{M^2_\beta+M^2_\gamma}{M^2_\alpha}
           +\frac{(M^2_\beta-M^2_\gamma)^2}{M^4_\alpha}}\,,
\label{three-mom}
\ee 
is the three momentum of the final state particles in the rest frame
of the decaying meson $\alpha$. A small correction due to vector meson
mixing~\ci{FKS2} is to be taken into account here; the $\phi - \omega$ 
mixing angle $\phi_V$ amounts to $3^\circ \pm 1^\circ$. The
PDG~\ci{PDG} average for the ratio \req{phi-ratio} leads to a value of 
$41.5^\circ \pm 2.2^\circ$ for the mixing angle. A new still
preliminary result on that ratio from KLOE~\ci{KLOE:04} provides an 
angle with an even smaller error namely $\phi=41.2^\circ \pm 1.1^\circ$. 
Both these values have not been used in the determination of the 
phenomenological mixing parameters quoted in Tab.\ \ref{tab:1}.  
Other reactions like the two-photon decays of the $\eta$ and $\etp$ 
\ci{feld:98} or the photon-meson transition form
factors~\ci{feld:98,passek} are sensitive to the decay constants. 
Thus, one finds for the two-photon decay width from PCAC~\ci{FKS1}
\ba
\Gamma[\eta\,\to \gamma\gamma]&=&\frac{\ale^2}{144\pi^3}\,M_\eta^3\,\,
           \left[\frac{5\sqrt{2}\,\cos{\phi}}{f_+} -
             \frac{\sin{\phi}}{f_s}\right]^2\,,\nn\\
\Gamma[\etp\to \gamma\gamma]&=&\frac{\ale^2}{144\pi^3}\,M_{\etp}^3\,
           \left[\frac{5\sqrt{2}\,\sin{\phi}}{f_+} +
             \frac{\cos{\phi}}{f_s}\right]^2\,,
\ea
and for the $P\gamma$ transition form  factor to next-to-leading order
(NLO) of perturbative QCD and leading-twist accuracy (in the $\overline{\rm
  MS}$ scheme)~\ci{passek}
\be
F_{P\gamma\gamma^*} \stackrel {{Q}^2 \to
\infty}{\longrightarrow} \frac{\sqrt{2} f^{\rm eff}_P}{Q^2} 
                               \Big[1-\frac53 \frac{\alpha_s}{\pi}\Big]\,,
\ee    
where the effective decay constants read ($f^{\rm eff}_\pi=f_\pi$)
\be
f^{\rm eff}_\eta\= (5f_+\cos{\phi}-\sqrt{2}f_s\sin{\phi})/3\,, \qquad
f^{\rm eff}_{\etp}\= (5f_+\sin{\phi}+\sqrt{2}f_s\cos{\phi})/3\,.
\ee
These processes fix the decay constants quite well~\ci{feld:98}.
In the form factor data there is a little hint at contributions from
the two-gluon Fock component of the $\etp$ which occurs to NLO
perturbative QCD~\ci{passek}.

The radiative decays of the $J/\psi$ into the $\eta$ and $\etp$
provide another interesting piece of information~\footnote{
These are examples of OZI-rule violating decays mediated by the
U$_{\rm A}$(1) anomaly.}. 
According to Novikov {\it et al}~\ci{nov} the photon is here emitted
from the charm quarks which subsequently annihilate into lighter quark
pairs through the effect of the U$_{\rm A}$(1) anomaly (see Fig.\
\ref{fig-decays}). This mechanism leads to the following result
for the ratio of the $\etp/\eta$ decay widths  
\be
\frac{\Gamma[J/\psi\to \gamma \etp]}{\Gamma[J/\psi\to \gamma \eta\,]}
 \= \left | \frac{\langle 0|\omega|\etp\rangle}
       {\langle 0|\omega|\eta\rangle}\right |^2\, 
     \left(\frac{k_{[J/\psi\,\gamma\,\etp]}}{k_{[J/\psi\,\gamma\,\eta\,]}}\right)^3\,.
\label{jpsi-ratio}
\ee   
Using Eqs.\ \req{matrix}, \req{mix-rel} and \req{eq:yz}, one can
express the ratio of the vacuum-particle matrix elements of $\omega$
by the mixing angle
\be
\frac{\langle 0|\omega|\etp\rangle}{\langle 0|\omega|\eta\,\rangle}\=
     \tan{\phi}\,\frac{M^2_{\etp}-M^2_{\pi^0}}{M^2_{\eta}-M^2_{\pi^0}}\,,
\label{ano-ratio}
\ee
a relation that has been derived by Ball et al \ci{ball} independently
on the quark-flavor mixing scheme. From the PDG~\ci{PDG} averages for 
the radiative $J/\Psi$ decays one obtains $\phi=39.0^\circ\pm 1.6^\circ$. 
 
The analysis of these reactions and a number of others leads to the 
set of phenomenological mixing parameters quoted in Tab.\ \ref{tab:1}. 
These values absorb corrections from higher orders of flavor symmetry 
breaking. As the comparison with the theoretical results reveals these 
effects are not large, they are on the $1 - 3\sigma$ level.

\section{The octet-singlet basis}
Transforming from the quark-flavor basis to the SU(3)${}_{\rm F}$
octet-singlet one by an appropiate orthogonal matrix, one easily
notices that the decay matrix in the octet-singlet basis has a more
complex structure than \req{decay} which is diagonal in the $\eta -
\etp$ sector. In addition to the state mixing angle 
$\theta = \phi-\theta_{\rm ideal}$ where the ideal mixing angle is 
$\arctan \sqrt{2}$, two more angles, $\theta_8$ and $\theta_1$, are 
needed in order to parameterize the weak decays of a physical meson 
through either the action of a singlet or an octet axial vector
current \ci{kaiser,feld:98}: 
\ba 
f^8_\eta\; &=& f_8 \cos{\theta_8}\,, \qquad f^1_\eta\, \= - f_1
\sin{\theta_1}\,,\nn\\
f^8_{\etp} &=& f_8 \sin{\theta_8}\,, \qquad f^1_{\etp} \= \phantom{-}f_1
\cos{\theta_1}\,.
\ea      
The mesons may also decay through strange and non-strange axial
vector currents. The corresponding decay constants can be
parameterized in a similar fashion
\ba 
f^+_\eta\, &=& f_+ \cos{\phi_+}\,, \qquad f^s_\eta\,\= - f_s
\sin{\phi_s}\,,\nn\\
f^+_{\etp} &=& f_+ \sin{\phi_+}\,, \qquad f^s_{\etp} \= \phantom{-}f_s
\cos{\phi_s}\,.
\ea   
In the quark-flavor mixing scheme described in the preceding sections,
the three angles are assumed fall together, $\phi=\phi_+=\phi_s$, and,
hence, the decay constants follow the pattern of state mixing. For the
sake of comparison I will keep the three different angles for the
remainder of this section. The two sets of decay constants are related 
to each other by
\ba
f_8 &=& \frac1{\sqrt{3}}\,\sqrt{f_+^2 + 2 f_s^2
    -2\sqrt{2}\,f_+f_s\,\sin{(\phi_+-\phi_s)}}\,, \nn\\
f_1 &=& \frac1{\sqrt{3}}\,\sqrt{2 f_+^2 +  f_s^2
     + 2\sqrt{2}\,f_+f_s\,\sin{(\phi_+-\phi_s)}}\,,
\label{f-rel}
\ea
and
\be
\tan{\theta_8}\=\frac{f_+\sin{\phi_+}-\sqrt{2} f_s\cos{\phi_s}}
                     {f_+\cos{\phi_+}+\sqrt{2} f_s\sin{\phi_s}}\,,
\qquad
\tan{\theta_1}\=\frac{f_s\sin{\phi_s}-\sqrt{2} f_+\cos{\phi_+}}
                     {f_s\cos{\phi_s}+\sqrt{2} f_+\sin{\phi_+}}\,.
\label{t-rel}
\ee
In the quark-flavor mixing scheme these relations simplify
drastically:
\be
\theta_8\= \phi - \arctan{(\sqrt{2} f_s/f_+)}\,,
\qquad
\theta_1\= \phi - \arctan{(\sqrt{2} f_+/f_s)}\,.
\ee
Depending on the strength of flavor symmetry breaking, embodied in
the ratio $f_+/f_s$, the mixing angles $\theta_8$ and $\theta_1$ may
differ from each other and from the state mixing angle, $\theta$,
substantially. The angle $\theta_8$ is smaller than $\theta_1$ if 
$f_s > f_+$. Only in the flavor symmetry limit, i.e.\ if $f_+=f_s$,
the three angles fall together. 

Evaluating the mixing parameters in the octet-singlet basis from the
theoretical and phenomenological sets of mixing parameters compiled in
Tab.\ \ref{tab:1}, one obtains the results shown in Tab.\ \ref{tab:2}.
Indeed the three mixing angles occuring in the octet-singlet basis, 
differ markedly. At the best mixing is simple in the quark-flavor
basis which is favored because of the smallness of those OZI-rule 
violations which are not induced by the U$_{\rm A}$(1) anomaly. On the 
other hand, SU(3)$_{\rm F}$ breaking is large and cannot be ignored. 
 
It should be clear from the above discussion that any mixing scheme 
which takes into account the U$_{\rm A}$(1) anomaly and includes the
proper masses of the physical mesons will lead to similar results for
the mixing parameters provided different values of the mixing angles, 
$\theta_8$, $\theta$ and $\theta_1$, are allowed for. This assertion is
indeed confirmed by the results found in many papers on this subject. 
For a detailed comparison of various mixing schemes it is refered to
the reviews~\ci{feld,uppsala}. With regard to the limited space
available for this work I refrain from repeating that and concentrate
on the comparison with recent work in which particular attention is
paid to the issue of the diverse mixing angles.
\begin{table}[h]
\renewcommand{\arraystretch}{1.3}
\tbl{Octet-singlet mixing parameters evaluated in the
  quark-flavor scheme from the parameters listed in Tab.\ \ref{tab:1}
  and comparison with other results. Errors for the latter results are
  not shown.}
{\begin{tabular}{ccccc| l}
$\theta$ & $\theta_8$ & $\theta_1$ & $f_8/f_\pi$ & $f_1/f_\pi$ & \\
\hline 
$-13.3^\circ$ & $-19.4^\circ$ & $-6.8^\circ$ & 1.19 & 1.10 &
theory~\ci{FKS1} \\
$-15.4^\circ$ & $-21.2^\circ$ & $-9.2^\circ$ & $1.26$ & $1.17$ & phen.~\ci{FKS1}\\
$\pm 1.0^\circ$ & $\pm 1.4^\circ$ & $\pm 1.4^\circ$ & $\pm 0.06$ & $\pm
0.04$ & \\\hline
 - & $-20.5^\circ$ & $-4^\circ$ & 1.28 & 1.10 & ChPT~\ci{kaiser}\\
$-10.5^\circ$ & $-20.0^\circ$ & $-1.5^\circ$ & 1.31 & 1.24 & ChPT~\ci{goity}\\ 
 - & $-10.8^\circ$ & $-15.6^\circ$ & $1.39$ & $1.39$ & sum rules~\ci{penn}\\
 - & $-23.8^\circ$ & $-2.4^\circ$ & $1.51$ & $1.29$ & phen.~\ci{frere}\\\hline
 \end{tabular}}
\label{tab:2}
\end{table}
\renewcommand{\arraystretch}{1.0}

Results from two NLO chiral perturbation theory analyses~\ci{kaiser,goity} 
are listed in Tab.\ \ref{tab:2} in which the $\etp$ meson is
consistently incorporated by means of the large $N_c$ limit. The
$\etp$ meson acts as a ninth Goldstone boson in this limit. For 
Ref.~\refcite{goity} where the experimental widths the  two-photon
decays into $\eta$ and $\etp$ are used as input in addition to the
masses and decay constants of the physical mesons, the averages of
their three fits results are quoted. In general fair agreement of the 
results is to be seen. Only the state mixing angle $\theta$, quoted 
in Ref.~\refcite{goity}, seems to be too small in magnitude. It
corresponds to $\phi=44.2^\circ$ with an uncertainty of about
$1.5^\circ$ which is rather large as compared with the above quoted 
values obtained from radiative $\phi$ and $J/\Psi$ decays. 

De Fazio and Pennington \ci{penn} calculated the decay constants
$f_P^{+,s}$ from QCD sum rules.  The octet-singlet mixing parameters, 
evaluated through Eqs.\ \req{f-rel} and \req{t-rel}, differ quite a
bit from the results obtained within the quark-flavor mixing scheme. 
The errors in the sum rule analysis are however large. The typically
amount to about $\pm 4^\circ$ for the mixing angles. Despite the large
error the result on $\theta_8$ is in conflict with the experimental 
value on the ratio \req{jpsi-ratio}. Since the ratio of the matrix 
elements \req{ano-ratio} can also be expressed as
\be
\frac{\langle 0|\omega|\etp\rangle}{\langle 0|\omega|\eta\,\rangle}\=
     -\cot{\theta_8}\,,
\label{ano-angle8}
\ee
experiment~\ci{PDG} leads to $|\theta_8|=22.0^\circ\pm 1.1^\circ$
directly.

Escribano and Fr\`ere ~\ci{frere} performed a phenomenological
analysis of a subset of the data exploited in Ref.~\refcite{FKS1},
namely $P\to \gamma\gamma$, $V\to \gamma P$ and $P\to \gamma V$, the
vector mesons include the $J/\Psi$. Their fits also provide clear 
evidence for substantially different values of $\theta_8$ and $\theta_1$, 
see Tab.\ \ref{tab:2}. The large value of $f_8$ (along with a large
$f_s$) they obtained, seems to be forced by the $\phi\to\gamma\etp$ 
width~\ci{PDG} which has not been included in the analyses of 
Refs.~\refcite{FKS1,kaiser,goity} and \refcite{penn}.

The angle $\theta_1$ exhibits the largest uncertainties of the mixing
parameters since none of the present data is really sensitive to
it. Helpful in pinning down its value would be more and better data on
the $\gamma - P$ transition form factors or the $g^{(*)}g^{(*)}\to P$ 
form factors~\ci{FKS2} which may for instance be obtained from the
process $pp\to {\rm jet}+ {\rm jet} +P$ in the central rapidity
region~\ci{frere:98}.

Inspection of Tab.\ \ref{tab:2} reveals strong flavor symmetry
breaking effects as characterized by the large values of
$|(\theta_8-\theta_1)/(\theta_8+\theta_1)|$. In contrast to this the
ratio $|(\phi_+-\phi_s)/(\phi_++\phi_s)|$ which can be evaluated from 
the information given in Tab.\ \ref{tab:2} with the help of the inverse 
of Eqs.\ \req{f-rel} and \req{t-rel}, is tiny; the values lie in the
range $0.01 - 0.04$ and, provided errors are available, are compatible
with zero~\footnote{
Admission of three mixing angles in the analysis
of the $P\gamma$ form factor leads to
$|(\phi_+-\phi_s)/(\phi_++\phi_s)|= 0.012\pm 0.026$~\ci{feld:98}.}. 
I.e.\, within the present accuracy of the data, the three mixing
angles $\phi_+$, $\phi$ and $\phi_s$ fall together. This observation 
strongly supports the central assumption \req{currents} of the 
quark-flavor mixing scheme. As the comparison with OZI-rule violating 
terms in the chiral Lagrangian reveals~\ci{feld} $\phi_+\simeq\phi\simeq\phi_s$
implies that OZI-rule violations except those mediated by the 
U$_{\rm A}$(1) anomaly, are negligible. It is important to realize in
this context that the validity of the OZI rule is a prerequisite for
the existence of process-independent mixing parameters.
\section{Isospin symmetry breaking}
Having discussed $\eta - \etp$ mixing in some detail and shown that
$\eta - \etp$ mixing is well understood, let me now turn
to ISB induced by $\pi^0 - \eta - \etp$ mixing. 
Defining the isospin-zero admixtures to the $\pi^0$ by
\be
|\pi^0\rangle \= |\eta_{\,-}\rangle + \eps\,|\eta\,\rangle + \eps^\prime
|\etp\rangle + {\cal O}(\lambda^2)\,,
\ee
one finds with the help of the matrix \req{matrix}
\ba
\eps &=& \psi + \beta\, \cos{\phi} = \cos{\phi}\, \left[ \frac12\, 
               \frac{m^2_{dd}-m^2_{uu}}{M^2_\eta - M^2_{\pi^0}} +
               z\right]\,,\nn\\
\eps^\prime &=&\quad\beta\, \sin{\phi}\quad = \sin{\phi}\, \left[\frac12\,
          \frac{m^2_{dd}-m^2_{uu}}{M_{\etp}^2 - M^2_{\pi^0}} +
               z\right]\,.
\label{angles}
\ea
The $f_u=f_d$ limits of these results, termed $\hat{\eps}=\eps(z=0)$ and
$\hat{\eps}^{\,\prime}=\eps^{\,\prime}(z=0)$ in the following,
coincide with the results reported in Ref.\ \refcite{FKS2}. The
numerical value of the quark mass term in \req{angles} may be
estimated from the $K^0-K^+$ mass difference corrected for mass
contributions of electromagnetic origin
\be
m_{dd}^2 - m_{uu}^2 \= 2\,\big[M^2_{K^0} - M^2_{K^+} - 
                               \Delta M^2_{K\,{\rm elm}}\big]\,.
\ee
According to Dashen's theorem \cite{dashen}, the electromagnetic
correction is given by~\footnote{
The QCD contribution to the $\pi^+ - \pi^0$ mass difference~\ci{gross}
is $\propto \eps$ and amounts to about 0.2 MeV which is much smaller
than the experimental value of 4.6 MeV.} 
\be
\Delta M^2_{K\,{\rm elm}}\= \big[M^2_{K^0}-M^2_{K^+}\big]_{\rm elm}=
                                 M^2_{\pi^0}-M^2_{\pi^+}\,,
\label{Dashen}
\ee
in the chiral limit and amounts to $-1.3\cdot 10^{-3}\, \gev^2$. 
This leads to $m^2_{dd}-m^2_{uu}=0.0104\gev^2$. However, finite quark
masses increase $\Delta M^2_{K\,{\rm elm}}$ substantially. The exact 
size of this enhancement is subject to controversy. Different authors
\ci{donoghue} obtained rather different values for the electromagnetic
mass splitting of the K mesons. For an estimate of the $z=0$ values of
the $\eta$ and $\etp$ admixtures to the $\pi^0$ I take the average of
the results for  $\Delta M^2_{K\,{\rm elm}}$ quoted in Ref.~\refcite{donoghue}
($(-2.4\pm 0.7)\cdot 10^{-3}\, \gev^2$). This way I obtain 
\be 
\hat{\eps}\= \eps(z=0) = 0.017 \pm 0.002\,, \qquad
\hat{\eps}^\prime\=\eps^\prime(z=0) = 0.004 \pm 0.001\,.
\label{numerics}
\ee
Due to the electromagnetic mass corrections the value for $\hat{\eps}$
is larger than that one quoted in Refs.~\refcite{FKS2,feld} and \refcite{uppsala}.

It is elucidating to turn the masses occuring in \req{angles} into quark
masses. With the help of the Gell-Mann-Oakes-Renner relations
\ci{GMOR} and the use of Dashen's theorem \req{Dashen}
one finds 
\be
\hat{\eps} = \sqrt{3}\, \cos{\phi}\, \eps_{\,0}\,,
\ee
with $\eps_{\,0}$ defined in \req{pcac}. The additional factor of
$\sqrt{3}\cos{\phi}$ ($=1.34$) would be unity if
$\phi=\theta_{\rm ideal}$, i.e.\ if the physical $\eta$ and $\etp$ mesons
are pure flavour octet and singlet states, respectively. In this case
there is no $\pi^0 - \etp$ mixing, i.e.\ $\hat{\eps}^\prime=0$. We now
see that the small value of $\eps_{\,0}$ ($= 0.011$) obtained by
Gross, Treiman and Wilczek~\ci{gross}, is a consequence  of the
disregard of $\eta-\etp$ mixing and the use of Dashen's result
\req{Dashen} for the electromagnetic contribution to the kaon mass
splitting.

Chao~\ci{chao:89} has also investigated $\pi^0 - \eta - \etp$ mixing on
exploiting the axial anomaly but he has used the PCAC hypothesis
instead of diagonalizing the mass matrix. Despite his assumption of
equal mixing angles, $\theta_1=\theta=\theta_8$, a supposition that has
been shown to be inadequate and theroretically inconsistent (see
Sect.\ 4), his results on $\eps$ and $\eps^\prime$ agree with the $z=0$
values~\req{numerics} within errors. NLO chiral perturbation theory in
the large $N_c$ limit~\ci{goity} leads to a slightly smaller value for 
the mixing angle ($\hat{\eps} \simeq 0.014 - 0.016$) than \req{numerics}. 

If the decay constants $f_u$ and $f_d$ differ from each other the
mixing angles may deviate from the values quoted in \req{numerics}
substantially. This potentially large effect is a source of considerable 
theoretical uncertainty of our understanding of ISB in the
pseudoscalar meson sector. In the absence of theoretical estimates for
$z$ one has to rely on phenomenological estimates of its value. Even
this difficult as will turn out in Sect.\ 7. 

\section{The vacuum-$\pi^0$ matrix element of the topological charge density} 
ISB as a consequence of $\pi^0-\eta-\etp$ mixing is accompanied by a
non-zero vacuum-$\pi^0$ matrix element of $\omega$. From the mixing
formulas derived in Sect.\ 2, one readily finds 
\be
\langle 0|\omega|\pi^0\rangle \=  \frac{\hat{\eps}}{\cos{\phi}}\,
               \frac{m^2_{ss} -M_{\pi^0}^2}{M_{\etp}^2-M_{\pi^0}^2}\,
                    \langle 0|\omega|\eta_{\,+}\rangle\,.
\label{piME}
\ee   
Any $z$ dependence cancels in this matrix element. As an immediate
consequence of \req{piME} the pion is contaminated by strange quarks
\be
\langle 0|2m_s \sb\, i\gamma_5\, s |\pi^0\rangle \= 
    - \frac{m^2_{ss}}{m_{ss}^2- M^2_{\pi^0}}\, \langle 0|\omega|\pi^0 \rangle\,. 
\label{sME}
\ee
This result implies a tiny violation of the OZI rule through the
anomaly. As a consequence the $\pi^0$ can decay through the strange
axial-vector current with a decay constant defined by $\langle
0|\partial^\mu J^s_{5\mu}|\pi^0\rangle=f_{\pi^0}^s M^2_{\pi^0}$. Using
the operator relation \req{anomaly} and Eqs.\ \req{piME}, \req{sME},
one obtains for this decay constant  
\be
f_{\pi^0}^s \= - \frac{\langle
  0|\omega|\pi^0\rangle}{m^2_{ss}-M^2_{\pi^0}}\,.
\ee  
It is very small, about $-5.2\cdot 10^{-3}\,f_\pi$, and will likely
have no experimental consequences.
For numerical estimates of the quantities just introduced one may 
use \req{constraint} and the mixing parameters quoted in the preceding
sections. 
  
The decays $\Psi(2S)\to J/\Psi P$ allow for an immediate test of
the prediction \req{piME}. As has been suggested in Ref.~\refcite{vol:80} 
these decays are dominated by a mechanism where the 
pseudoscalar meson is coupled to the $c\bar{c}$ system through the
effect of the anomaly. Hence, the $\pi^0/\eta$ ratio of these
processes is given by
\be 
\frac{\Gamma[\Psi(2S)\to J/\Psi\, \pi^0]}{\Gamma[\Psi(2S)\to J/\Psi\,\eta\,]}
\= \left| \frac{\langle 0|\omega|\pi^0\rangle}{\langle
  0|\omega|\eta\,\rangle}\right|^2\;
  \left(\frac{k_{[\Psi(2S),J/\Psi,\pi^0]}}{k_{[\Psi(2S),J/\Psi,\eta]}}\right)^3\,, 
\label{ratio-psip}
\ee
in analogy to \req{jpsi-ratio}. The ratio of the gluonic matrix elements
follows from \req{piME} and results presented in Sect.\ 2. It reads
\be
\frac{\langle 0|\omega|\pi^0\rangle}{\langle
  0|\omega|\eta\,\rangle} \= \frac{\hat{\eps}}{\cos^2{\phi}}\,. 
\label{ratio-ano}
\ee
It is to be stressed that Eq.\ \req{ratio-ano} includes $\pi^0 - \etp$
mixing. It is responsible for the factor $1/\cos^2{\phi}$ which amounts
to about 1.7. This is so because the small mixing angle
$\eps^\prime$ is enhanced by the large matrix element $\langle
0|\omega|\etp\rangle$, see \req{ano-angle8}. The importance of $\pi^0
- \etp$ mixing in Eq.\ \req{ratio-psip} has been pointed out by
Genz~\ci{genz} long time ago. Using the recent, rather accurate BES
measurement~\ci{BES04} of the ratio \req{ratio-psip}, one extracts the
mixing angle
\be
\hat {\eps}\,[\Psi(2S)\to J/\Psi P] \= 0.030 \pm 0.002\,, 
\ee  
which is large as compared to the theoretical estimate \req{numerics}. 
The origin of this  discrepancy is not understood. Neglected
electromagnetic contributions to the transitions $\Psi(2S)\to J/\Psi
\,P$ are likely not the reason~\ci{don-wyl}. A possible explanation
could be that the coupling of the pseudoscalar mesons to the
$c\bar{c}$ system is not or only partially controlled by the anomaly. 
This possibility has been suggested in Ref.\ \refcite{casa} and discussed 
whithin the framework of the heavy quark effective theory. Since the
strength of this new mechanism has not been predicted in Ref.\
\refcite{casa} its bearing on the determination of $\hat{\eps}$ cannot 
be estimated. A better understanding of the $\Psi(2S)\to J/\Psi P$
decay may also affect the determination of quark masses~\ci{leut:96a}.  
    
One may wonder whether the isospin-violating radiative decay of the
$J/\Psi$ into the $\pi^0$ is also under control of the U$_A(1)$
anomaly. This is, however, not the case in contrast to the $\gamma
\eta$ and $\gamma\etp$ channels which are well described by the
anomaly contribution. An estimate of the $J/\Psi\to\gamma\pi^0$ decay
width using \req{ratio-ano} and the analogue of \req{jpsi-ratio}, 
provides values that are too small by more than an order of
magnitude. In fact the radiative $J/\Psi$ transition into the 
$\pi^0$ is mediated by the higher order electromagnetic process 
$c\cb\to \gamma\to \pi^0 \gamma$ and by a vector meson dominance
contribution $J/\Psi\to\rho \pi^0\to \gamma\pi^0$.

\section{Phenomenological determination of the mixing angles $\eps$} 
In this section ISB-violating processes which are not under control
of particle-vacuum matrix elements of $\omega$, will be discussed. 
The full mixing angle $\eps$ and not 
just its $z=0$ value, occurs under these circumstances. Care is to be 
taken that only ISB within QCD are analyzed, eventual electromagnetic 
effects have to be subtracted. 

Clear signals for ISB and/or charge symmetry breaking have been
observed in a number of hadronic reactions. The extraction of the
mixing angle from these data is however difficult and model
dependent. The ratio of the $\pi^+d\to pp\eta$ and $\pi^-d\to nn \eta$ 
deviates from unity, the charge symmetry result, 
experimentally~\ci{tippens:01}. On the basis of a rather simple model
that includes $\pi^0 - \eta$ state mixing and a number of corrections
which take into accounts effects such as differences in the meson-nucleon
coupling constants or the proton-neutron mass difference but ignores
mixing with the $\etp$, the authors of Ref.~\refcite{tippens:01} extracted
a $\pi^0 -\eta$ mixing angle of 
\be
\eps\,[\pi^{\pm} d\to NN \eta] \= 0.026 \pm 0.007\,,
\label{eps(pid)}
\ee      
from their data. The non-zero forward-backward asymmetry in $np\to d\pi^0$
measured at TRIUMF~\ci{opper:03} is in conflict with charge symmetry. 
The phenomenological analysis of this data suffers from large
ambiguities. A combination of contributions from $\pi^0-\eta$ mixing, 
from the $u-d$ quark-mass difference and from electromagnetic effects 
in the nucleon-mass difference, controls the asymmetry~\ci{gardestig:04}.  
A further complication arises from the fact that the mixing
contribution is in fact given by a product of the $\pi^0-\eta$ mixing
angle and the badly known $\eta$-nucleon coupling constant~\ci{dumbrajs}. 
This complicated situation prevents an extraction of the mixing angle
with a significant accuracy. But the measured asymmetry in $np\to d\pi^0$ 
is compatible with the present knowledge of the various quantities
occuring in the theoretical result. The cross section data~\ci{steph:03} for 
$dd\to {}^4He\,\pi^0$ have not yet completely been analyzed 
theoretically~\ci{garde} while the COSY measurement of the ratio of
the $pd\to {}^3H\, \pi^+$ and $pd\to {}^3He\, \pi^0$ 
cross sections provides only a very weak signal for ISB \ci{abdel:03}.
More precise data on the latter cross sections are required before   
one can draw a definite conclusion here.

The $\eta$ and $\etp$ decays into $3\pi$ violate G-parity and hence
isospin symmetry. Since electromagnetic contributions are strongly 
suppressed~\ci{suther} ISB is here mainly of QCD origin and can be 
estimated through $\pi^0 - \eta - \etp$ mixing. Recent analyses of 
the $\eta$ decays within NLO chiral perturbation 
theory~\ci{kambor,gasser-uppsala} can be summarized as
\be 
\Gamma[\eta\to \pi^+\pi^-\pi^0] \=
                      \left(\frac{\eps}{\eps_0}\right)^2\, \overline{\Gamma}\,,
\label{eta-decay}
\ee
where $\bar{\Gamma}\simeq 180 - 230\,\ev$. This result matches the 
experimental value of $292\pm 17\,\ev$ if 
\be
\eps\,[\eta\;{\rm decay}] =0.014 \pm 0.001\,.  
\label{eps(eta)}
\ee
This value corresponds to the result obtained within the quark-flavor
mixing scheme provided $\eta - \etp$ mixing is ignored. In fact 
\req{eps(eta)} is a typical result of NLO chiral perturbation theory 
in which the $\etp$ is not considered as an explicit degree of freedom 
in the effective Lagrangian. Its effect is, however, partially
embodied in the effective coupling constants of the theory. The NLO 
terms of chiral perturbation theory whith which a number of low energy
processes can be calculated rather precisely, comprise contributions
to ISB of various origin. Hence, the comparison with results from the
quark-flavor mixing scheme which allows for an investigation of many 
processes at low and high energies, is not straightforward.

The ratio of the $\etp\to 3\pi^0$ and $\etp\to \eta\pi^0\pi^0$ decay 
widths can be written as~\ci{gross}
\be
\frac{\Gamma[\etp\to 3\pi^0]}{\Gamma[\etp\to\pi^0\pi^0\eta ]} \= R\,
     \left(\frac{\eps}{\eps_0}\right)^2\,,
\ee
if electromagnetic contributions can be neglected. Coon {\it et al}
\ci{coon} estimated the factor $R$ relying on PCAC ideas and
taking into account the experimentally observed mild dependence of the
process amplitudes on kinematics~\ci{alde}. Their result on $R$
combined with the experimental value of $(7.5\pm 1.3)\cdot 10^{-2}$
for the ratio of the decay widths~\ci{PDG} leads to 
\be
\eps\,[\etp\;{\rm decay}] \= 0.021\pm 0.002\,.  
\label{eps(etap)}
\ee
More accurate experimental data and a revision of the theoretical
analysis is advisable. A calculation of the $\etp$ decays within the
framework of chiral perturbation theory is not available.

There are several $J/\Psi$ decays into vector and pseudoscalar mesons
which violate isospin symmetry, e.g.\ $J/\Psi\to \rho^0\eta, \omega\pi^0,
\phi\pi^0$. As compared to the corresponding isospin allowed decays,
$J/\Psi\to \rho^0\pi^0, \omega\eta,\phi\eta$, they are typically
suppressed in experiment by a factor of about $10^{-1} - 10^{-2}$ in
accord with expectations for an electromagnetic decay mechanism. 
Contributions from ISB mechanisms within QCD are negligible small. 
Indeed, within the $\pi^0-\eta-\etp$ mixing scenario, ratios like 
$\Gamma[J/\Psi\to \phi\pi^0]/\Gamma[J/\Psi\to \phi\eta]$ would be
proportional to $\epsilon^2$ and would therefore amount to only $10^{-4} -
10^{-3}$. As is the case for the process $J/\Psi\to \gamma\pi^0$ the
contribution from $\pi^0-\eta$ mixing to the radiative decay of the
$\phi$ meson into the $\pi^0$ is negligible small; it involves the 
$\eta_s$ component of the $\pi^0$ (see Fig.\ \ref{fig-decays}) which 
is proportional to $\eps$. The process $\phi\to\gamma\pi^0$ proceeds 
through mechanisms similar to those occuring in $J/\Psi\to\gamma\pi^0$.  

Recently a new meson, the $D_{sJ}^*(2317)$, has been 
observed~\ci{BABAR:03,BELLE,besson}. Its experimental properties are
consistent with a $J^P=0^+$ $c\bar{s}$ state interpretation. 
It decays into $D_s^+\pi^0$, no other decay channel have been observed
as yet. This ISB decay likely proceeds through the production of the
$\eta_s$ component of the pion and consequently its width is $\propto
\hat{\epsilon}^2$, see \req{matrix}, \req{iso-rel}. Estimates of the
decay width~\ci{cho} using chiral perturbation theory and the heavy
quark effective theory provides values way below the present experimental
upper bound for the total $D_{sJ}^*(2317)$ width. A second new meson 
which also decays into the isospin symmetry violating channel 
$D_{sJ}^+(2457)\to D_s^{*+}\pi^0$ has been observed too~\ci{BELLE,besson}. 
Once the decay widths of these two processes will be measured more 
precisely the mixing picture of ISB can again be probed.
 
Isospin violations within QCD are also of relevance in CP-violating
processes whose analyses are intricate and do not allow a simple
extraction of the mixing angles. An example is set by the 
$K^0\to \pi\pi$ decays~\ci{holstein}. In a recent 
analysis~\ci{neufeld} of this process whithin NLO chiral perturbation 
theory the value \req{eps(eta)} for the mixing angle $\eps$ has been
used. The above comments on the comparison of results from this theory 
with others apply here as well. Another example of ISB effects in weak 
decays is the process $B\to \pi\pi$. ISB spoils the 
triangle relation obeyed by the amplitudes for the three processes
$B^+\to\pi^+\pi^0$, $B^0\to \pi^0\pi^0$ and $B^0\to \pi^+\pi^-$ in the
isospin symmetry limit and consequently affects the determination of
the CKM angle 
$\alpha=\arg{[-V^{\phantom{*}}_{td}V^*_{tb}/(V^{\phantom{*}}_{ud} V^*_{ub})]}$ 
\ci{gardner,gronau}. As estimated by Gardner~\ci{gardner} the mixing 
angles \req{numerics} lead to a correction of $\simeq 4^\circ$ which 
is substantial but still smaller than the error of the present
experimental result: $\alpha=100^\circ \pm 13^\circ$ \ci{hep-ex}.

\section{Summary}
A detailed theoretical and phenomenological analysis showed that the
quark-flavor mixing scheme provides a consistent description
of $\eta - \etp$ mixing. On exploiting the divergencies
of the axial vector currents all basis mixing parameters can be
determined for given masses of the physical  mesons.  It turned out
that flavor symmetry breaking manifests itself differently in the
mixing properties of states and decay constants, their
parameterization requires three different mixing angles in general.
Only in the quark-flavor basis the three angles fall together
approximately and mixing is simple. Other approaches to $\eta - \etp$
mixing lead to similar results provided the U$_{\rm A}$(1) anomaly and
the masses of the physical mesons are taken into account.

Inclusion of the $\pi^0$ into the quark-flavor mixing scheme induces
ISB of about $2\%$ on the amplitude level. Extraction of the $\pi^0 -
\eta$ mixing angle from experiment, on the other hand, provide
values in the range 0.02 - 0.03. Thus, the exact magnitude of ISB
through $\pi^0 -\eta$ mixing, is not yet determined and more work is
clearly needed. There are other sources of ISB within QCD besides
$\pi^0 -\eta$ mixing. Their estimate is however difficult and often
not or only partially taken into account in analyses. One of these
sources of ISB is a possible difference between the basic decay 
constants $f_u$ and $f_d$ or, respectively, the parameter $z$ in 
Eq.~\req{eq:yz}. From the results \req{eps(pid)}, \req{eps(eta)}, 
\req{eps(etap)} there is no clear evidence for a non-zero value of 
$z$. One may only conclude that $z$ is smaller than about 0.015. 
This entails a difference between $f_u$ and $f_d$ of less than $3\%$.   

{\it Acknowledgements} It is a pleasure to thank A.~Bernstein, F.~De Fazio,
R.~Escribano, W.~Kluge, G.~Nardulli and H.~Neufeld for discussions.

\end{document}